\documentclass[11pt]{article}

\usepackage[english]{babel}
\usepackage[letterpaper,top=1in,bottom=1in,left=1in,right=1in,marginparwidth=0.75in]{geometry}

\usepackage{lineno}
\usepackage[version=4]{mhchem} 
\usepackage{siunitx}
\usepackage{graphicx}
\usepackage{setspace}

\title{Structured Illumination for Surface-Resolved \\ Grazing-Incidence X-ray Scattering}

\author{Do\u ga G\" ursoy$^1$\footnote{E-mail: dgursoy@anl.gov}, Xiaogang Yang$^2$, Dina Sheyfer$^1$, Michael Wojcik$^1$, Ruipeng Li$^2$, Esther Tsai$^3$ \\ \\
$^1$ Advanced Photon Source, Argonne National Laboratory \\
$^2$ National Synchrotron Light Source II, Brookehaven National Laboratory \\
$^3$ Center for Functional Nanomaterials, Brookehaven National Laboratory}

\begin{document}
\doublespacing
\maketitle

\begin{abstract}
We present a computational imaging technique for imaging thin films at grazing-incidence (GI) angles by incorporating structured illumination into existing GI X-ray scattering setups. This method involves scanning a micro-coded aperture across the incident X-ray beam at a grazing angle, followed by computational reconstruction to extract localized structural and scattering information along the beam footprint on the sample. Unlike conventional GI X-ray scattering methods, which provide only averaged structural data, our approach offers localized scattering information. We detail the underlying principles of this technique and demonstrate its effectiveness through experimental results on an organic semiconductor thin film.
\end{abstract}

\section{Introduction}

X-ray scattering methods are widely employed for material characterization in a variety of fields, including chemistry, physics, materials science, and nanotechnology. These techniques are instrumental in probing the structural properties of different materials, providing critical insights into their morphology, phase, and crystalline properties. Among these, grazing-incidence (GI) X-ray scattering techniques, such as grazing-incidence wide-angle X-ray scattering (GIWAXS) and grazing-incidence small-angle X-ray scattering (GISAXS), are particularly effective for characterizing thin films, as they probe surface and near-surface regions without substrate limitations. These methods offer statistical information on crystalline domains and morphology, making them highly effective for investigating complex materials, including polymer blends, organic semiconductors, and self-assembled nanostructures. In particular, GIWAXS is used as a non-destructive tool for analyzing both in-plane and out-of-plane features, and plays a key role in monitoring real-time thin-film growth processes, offering crucial insights for optimizing materials in applications such as solar cells, transistors, and coatings \cite{lim1987grazing, pietsch2004high, stribeck2007x, zhang2017stable, muller2014active, benvenuti2023beyond, richter2017morphology, sidhik2024two}.

Despite its versatility, conventional GI X-ray scattering methods face challenges in resolving individual domains, e.g. variation in orientation or crystallinity within or across domains, due to the extended X-ray footprint in grazing incidence geometry. To overcome these limitations, integrating computed tomography (CT) with GISAXS or GIWAXS has been proposed to reconstruct local variations in material properties \cite{ogawa2015visualizing, ogawa2020improving, ogawa2017visualization} and domain orientations \cite{tsai2021grazing}. However, CT-based approaches present practical challenges, particularly the requirement for sample rotation, which is susceptible to several issues, including runout errors, stage drift, and alignment difficulties, especially at tilted geometries. These complications make data collection and reconstruction challenging. Moreover, incomplete angular data, due to the limited visibility of scattering peaks, result in artifacts that degrade resolution. The need for complete data stacks further restricts the feasibility of real-time visualization. Beamline interruptions, such as beam dumps, exacerbate the difficulty of collecting comprehensive datasets.

\begin{figure}
\begin{center}
\includegraphics{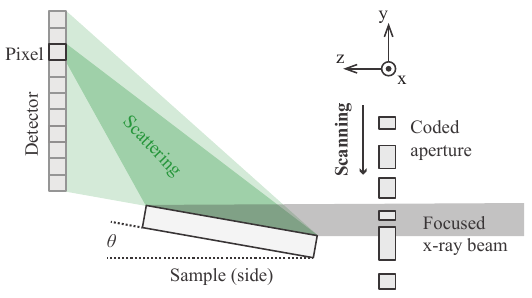}
\caption{Schematic of the SI-GID setup: A focused x-ray beam illuminates a sample at a grazing-incidence angle, with scattered signals recorded by a 2D pixel detector. A coded aperture structures the beam, and scanning encodes scattering signals for each pixel.}
\end{center}
\label{fig1}
\end{figure}

To overcome these challenges, we propose a structured illumination technique \cite{forbes2021structured} for grazing-incidence diffraction (SI-GID), which resolves local structural variations without requiring sample rotation. This method is compatible with existing sample environments and in-situ experiments, providing a robust alternative to conventional approaches. As illustrated in Figure~\ref{fig1}, our SI-GID approach utilizes absorbing micro-coded apertures to structure the X-ray beam, which is then scanned vertically along the y-axis to introduce measurement diversity. In this setup, for a beam with vertical and horizontal dimensions $D_v$ and $D_h$, respectively, the beam footprint extends along the z-direction as $D_v/\sin\theta$ due to the small incidence angle. Meanwhile, the resolution along the x-axis is determined by the horizontal beam width, $D_h$. Scanning the aperture along y-axis mimics the effect of sample rotation in traditional tomography, enabling surface spatial resolution without the associated instabilities. By keeping the sample stationary, SI-GID eliminates issues such as runout errors, stage drift, and alignment challenges. The fixed configuration also simplifies calibration, enabling rapid determination of z-axis resolution without requiring full tomographic (x-axis scans and rotational) data, facilitating time-resolved, high-resolution imaging. 

\section{Methods}

\subsection{Working Principle}

The underlying principle of this approach involves modulating the X-rays incident on the sample surface using a micro-coded aperture to generate shadowing patterns. Mathematically, let the unknown surface structure we aim to resolve be denoted as $\boldsymbol{s} = [s_1, \dots, s_N]^T$, and the aperture structure at position \(p_1\) be represented by $\boldsymbol{a}_{p_1} = [a_{p_1}, \dots, a_{p_1+N-1}]^T$, where the coefficients indicate the transmissivity at each location of the coded aperture. Here, $N$ represents the number of discrete segments in the piecewise-constant representation of the entire aperture. We can then formulate the following system of equations: 
\begin{equation}
\label{eq1}
\begin{bmatrix}
    d_{1} \\
    \vdots \\
    d_{M}
\end{bmatrix}
=
\begin{bmatrix}
    a_{p_1} & \dots  & a_{p_1+N-1} \\
    \vdots &  \ddots & \vdots \\
    a_{p_M} & \dots  & a_{p_M+N-1}
\end{bmatrix}
\begin{bmatrix}
    s_{1} \\
    \vdots \\
    s_{N}
\end{bmatrix},
\end{equation}
where $\boldsymbol{d} = [d_1, \dots, d_M]^T$ represents the intensity vector obtained from a single detector pixel after translating the coded mask $M$ times. The positions of the coded aperture are tracked by the vector $\boldsymbol{p} = [p_1, \dots, p_N]^T$, which are determined by the encoders of the linear motor during the aperture scanning process. This system can be compactly expressed in matrix form as:
\begin{equation}
\label{eq2}
    \boldsymbol{d} = \boldsymbol{A} \boldsymbol{s},
\end{equation}
where \(\boldsymbol{A}\) is the coding matrix with dimensions \(M \times N\).

In fundamental terms, the image reconstruction problem can be formulated as a least-squares optimization. Given the measured intensity data vector $\boldsymbol{d}$ and the system matrix $\boldsymbol{A}$, which represents the linear relationship between the unknown signal vector $\boldsymbol{s}$ and the measured data, the problem is expressed as:
\begin{equation}
\label{eq3}
\min_{\boldsymbol{s}} \frac{1}{2} \| \boldsymbol{d} - \boldsymbol{A} \boldsymbol{s} \|_2^2
\end{equation}
where $\| \cdot \|_2^2$ denotes the Euclidean norm. The matrix $\boldsymbol{A}$ encapsulates the relationship between the coded aperture pattern and the scattered x-ray signals into a pixel. By solving this problem for each pixel in the detector, the scattered signal on the surface along the beam footprint can be resolved spatially. Note that the resolution of the reconstructed image is independent of the discretization parameter $N$. Instead, it is determined by the rank of $\boldsymbol{A}$, the problem formulation in Equation~\ref{eq3}, and the method used to solve it.

\subsection{Experiment Setup}

\begin{figure}
\begin{center}
\includegraphics{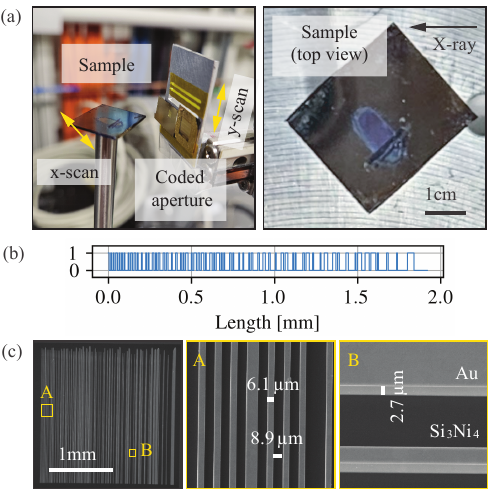}
\caption{(a) Schematic of the setup showing the coded aperture, sample, and detector, with a zoomed view of the sample and aperture. (b) Top-down optical photograph of the sample. (c) 256-bit coded aperture design. (d) SEM images of the fabricated aperture with zoomed-in views of regions A and B.}
\end{center}
\label{fig2}
\end{figure}

We validated the structured illumination grazing-incidence diffraction (SI-GID) technique through 2D structural imaging of organic semiconductor thin films, a model system that highlights the importance of understanding the structure-property-performance relationship in functional materials. In organic electronics, particularly in the design of transistors, correlating thin-film structures with device performance is critical for optimization \cite{giri2014one, lee2012guiding}. Conventional X-ray scattering techniques, while effective for providing statistical information on lattice structures, often fail to capture localized details such as domain shape, orientation, and polymorphism, which are essential for advancing material design.

The organic films were prepared by casting a 1:1 mixture of C8- and C12-BTBT ([1]benzothieno[3,2-b][1]benzothiophene) in the melt state and cooling at \SI{0.1}{\degree/\minute}, formed millimeter-sized domains on silicon substrates pre-coated with a \SI{100}{\nano\meter} polymethyl methacrylate layer. These films exhibited a uniform out-of-plane orientation across domains, while in-plane orientation varied significantly between them. This variation in in-plane orientation and polymorphism near localized domain boundaries is particularly critical for understanding and optimizing device performance.

Experiments were conducted at the Complex Materials Scattering (CMS) beamline (11-BM) at the National Synchrotron Light Source II, using a grazing-incidence wide-angle X-ray scattering (GIWAXS) setup with a coded aperture for structured illumination. The general layout of the setup is illustrated in Figure~\ref{fig2}. A \SI{15}{\kilo\electronvolt} X-ray beam of \SI{200}{\micro\meter} size illuminated the sample at a \SI{0.3}{\degree} grazing angle, leading to a surface illumination of approximately \SI{1.9}{\centi\meter} by \SI{200}{\micro\meter} on the sample. A coded aperture with a de Bruijn sequence of 256 bits, fabricated via direct-write lithography, introduced measurement diversity. Positioned \SI{30}{\milli\meter} upstream of the sample, the aperture was scanned vertically over \SI{600}{\micro\meter} in \SI{1}{\micro\meter} steps, achieving a z-axis resolution of ~\SI{190}{\micro\meter}.

\begin{figure}
\begin{center}
\includegraphics[width=0.5\textwidth]{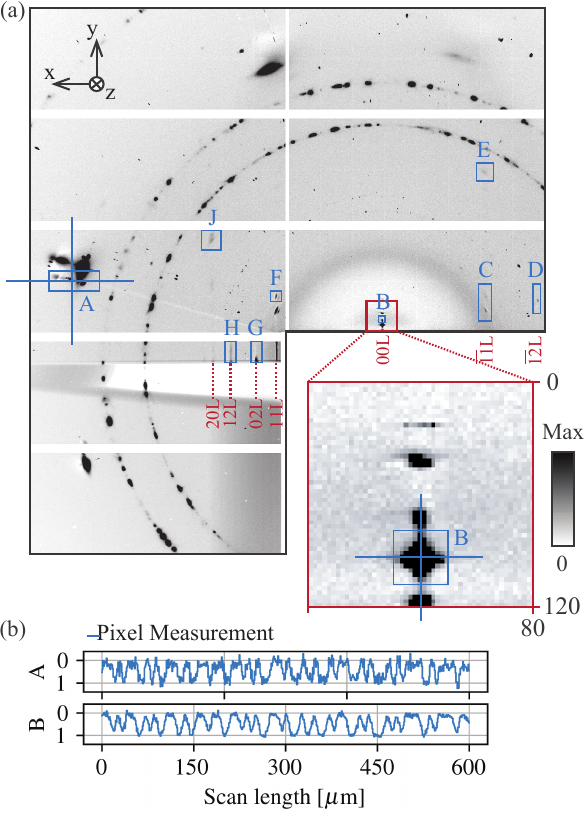}
\caption{(a) Representative image (average of collected frames) recorded using a Pilatus800k detector, showing a zoomed-in view of the scattered signals from the sample. The analyzed peaks are labeled in blue, ranging from A to J. (b) Measured intensities corresponding to pixels A and B obtained by scanning the coded aperture over a range of \SI{600}{\micro\meter}.}
\end{center}
\label{fig3}
\end{figure}

The scattered X-rays were captured using a Pilatus800k detector with a pixel size of \SI{172}{\micro\meter}, positioned \SI{0.17}{\meter} meters downstream of the sample. A representative intensity recording obtained from this setup is shown in Figure~\ref{fig3}, which features zoomed-in sections of the scattered signals from two pixels, A and B. Pixel A corresponds to signals scattered from the sample, while Pixel B detects signals from the silicon substrate beneath the sample. The corresponding measurement data for both pixels, as a function of the aperture scan, is presented in Figure~\ref{fig3}b. These data reveal intensity modulations induced by the coded aperture scanning, which align with the encoded structure of the aperture, demonstrating the effectiveness of the structured illumination technique in capturing spatially resolved scattering information.

To obtain 2D surface information, horizontal scans of the sample were performed in \SI{200}{\micro\meter} steps, matching the beam spot size, over a \SI{4}{\milli\meter} range, resulting in 26 data points to reconstruct surface images. Data were collected with an exposure time of \SI{1}{\second} per scan, synchronized with the stepping motion of the coded aperture. Digital reconstruction utilized nonnegative least squares optimization \cite{lawson1995solving}, implemented in SciPy \cite{virtanen2020scipy}, to resolve scattering data across the beam footprint, enabling high-resolution spatial mapping of the sample's surface structure.

\section{Results}

\begin{figure*}
\begin{center}
\includegraphics[width=\textwidth]{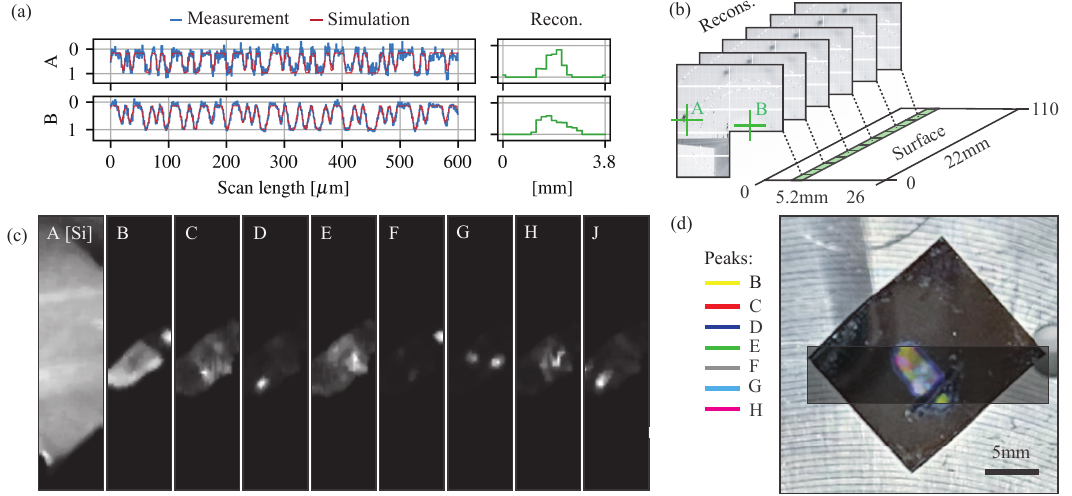}
\caption{(a) Reconstructed signals for pixels A and B, compared with simulations and experimental data, showing good agreement. (b) Surface-resolved scattering images across all detector pixels, enabling depth-specific visualization. (c) Reconstructed surface images for peaks identified in Fig.~\ref{fig3}. (d) Overlay of selected reconstructed peaks on the optical image, highlighting spatial alignment.}
\end{center}
\label{fig4}
\end{figure*}

The reconstruction results, presented in Figure~\ref{fig4}, provide x-ray scattering images that represent the sample's surface as a function of the x and z positions for selected peaks resolved on the surface. To illustrate the process, we present the reconstructed signal from a single pixel readout in Figure~\ref{fig4}(a) alongside the corresponding simulated signal, demonstrating good agreement with the measurements. The reconstruction process was then repeated for all detector pixels, resulting in surface-resolved scattering images, as shown in Figure~\ref{fig4}(b). This approach is powerful as it enables the retrieval of 2D scattering pattern for each pixel on the sample, allowing for the study of thin film inhomogeneity or grain boundaries. Local thin film structures can provide insights or design principles for functional material properties or performance. 

To further explore structural variations, we selected several regions of interest (ROIs) with key high-intensity scattering peaks identified in Figure~\ref{fig3} and depicted their corresponding reconstructions in Figure~\ref{fig4}(c). This approach effectively resolves both in-plane and out-of-plane structural variations associated with the targeted peaks. The spatial distribution of these peaks is shown in Figure~\ref{fig4}(c), demonstrating strong alignment with the optical photograph of the sample presented in Figure~\ref{fig4}(d). Additionally, the matching features from the silicon substrate between the optical and reconstructed images provide convincing validation of the outcome, further supporting the reliability of the method.

\section{Discussions}

We have presented SI-GID as a structured illumination method for GIWAXS that enables rapid surface and near-surface imaging of planar samples without requiring sample rotation. Conventional GI X-ray scattering methods typically provide structural or morphological information averaged over large sample areas due to the extended beam footprint. While tomographic methods combined with GIWAXS can provide localized information~\cite{tsai2021grazing}, it assumes uniform lattice structure, precise rotational alignment, and complete tomographic datasets. Insufficient angular sampling or missing angles in tomography introduce artifacts and compromise resolution.

SI-GID overcomes the limitations of conventional methods by employing a scanning coded aperture to generate structured illumination, allowing direct reconstruction of localized structural information. For each pixel on the sample, scattering patterns containing structural or morphological details are reconstructed, enabling the study of inhomogeneity and polymorphism. This pixel-level information on crystalline structure, composition, and orientation elucidates structure-performance relationships and supports in-situ, time-resolved experiments. Materials previously studied with standard grazing-incidence methods, such as perovskites for solar panels, can benefit significantly from SI-GID. For instance, resolving micron-sized domain orientations provides valuable insights for designing efficient solar cells~\cite{li2022light, sidhik2024two}.

By eliminating the need for a rotation stage, SI-GID is highly adaptable to existing sample environments or custom setups, inherently improving measurement stability by avoiding rotational drift and vibration. Additionally, it can support in-situ, time-resolved experiments, where a full rotation of the sample can be challenging to achieve. We demonstrated the method’s effectiveness through the surface reconstruction of organic thin films, revealing significant local structural features. While SI-GID may miss crystalline domains that do not satisfy the Bragg condition, combining it with few-angle tomography or small-range rocking scans can address this limitation for comprehensive domain orientation analysis.

\section{Conclusion}

In conclusion, SI-GID reveals localized variations in crystalline orientation and phase composition, offering detailed insights into material inhomogeneity and polymorphism. Its stability, compatibility with existing sample environments, and ability to provide spatially resolved scattering information make it a versatile and robust alternative to conventional methods. By enabling real-time, in-situ studies without the need for sample rotation, SI-GID has the potential to become a capable tool for advancing the understanding of thin films and related materials.

\section*{Acknowledgments} 

This research used resources of the Advanced Photon Source at Argonne National Laboratory and beamline 11BM (CMS) at the National Synchrotron Light Source II, supported by the U.S. DOE Office of Science under Contracts DE-AC02-06CH11357 and DE-SC0012704. X-ray scattering was conducted via the partner user program at the Center for Functional Nanomaterials, also operated by Brookhaven National Laboratory.

\bibliographystyle{ieeetr}
\bibliography{refs}

\end{document}